\documentclass[onecolumn,english,smallextended]{svjour3}
\usepackage{amsmath}
\usepackage{graphicx}
\usepackage{amssymb}

\usepackage{babel}

\begin{document}

\title{Gravitational sources induced by exotic smoothness and fermions as knot complements}

\author{Torsten Asselmeyer-Maluga \and Carl H. Brans}

\institute{T. Asselmeyer-Maluga \at Aerospace Center (DLR), Berlin \\ \email{torsten.asselmeyer-maluga@dlr.de}
\and C.H. Brans \at Loyola University, New Orleans\\ http://www.loyno.edu/~brans \\ \email{brans@loyno.edu}
}

\date{Received: date / Accepted: date}
\maketitle
\begin{abstract}
In this paper we will discuss Brans conjecture that exotic smoothness serves as an additional gravitational source naturally arising from the handlebody construction of the exotic $\mathbb{R}^{4}$. We will consider the two possible classes, the large and the small exotic $\mathbb{R}^{4}$. Then we calculate the Einstein-Hilbert action for both exotic $\mathbb{R}^{4}$ to show the apearance of spinor fields. Then we discuss the physical properties of these spinor fields to relate them to fermions. Finally we identify the corresponding 3-manifolds as knot complements of hyperbolic knots, i.e. the knot complements are hyperbolic 3-manifolds with finite volume. With the help of this result we confirm the Brans conjecture for both kinds of exotic $\mathbb{R}^{4}$. 
\end{abstract}

\section{Introduction}

The existence of exotic (non-standard) smoothness on topologically
simple 4-manifolds such as exotic $\mathbb{R}^{4}$ or $S^{3}\times\mathbb{R},$
has been known since the early eighties but the use of them in physical
theories has been seriously hampered by the absence of finite coordinate
presentations. However, the work of Bizaca and Gompf \cite{BizGom:96}
provides a handle body representation of an exotic $\mathbb{R}^{4}$
which can serve as an infinite, but periodic, coordinate representation.

Thus we are looking for the decomposition of manifolds into small
non-trivial, easily controlled objects (like handles). As an example
consider the 2-torus $T^{2}=S^{1}\times S^{1}$ usually covered by
at least 4 charts. However, it can be also decomposed using two 1-handles
$D^{1}\times D^{1}$ attached to the $0-$handle $D^{0}\times D^{2}=D^{2}$
along their boundary $\partial D^{2}=S^{1}$ via the boundary component
of the 1-handle $\partial D^{1}\times D^{1}=S^{0}\times D^{1}$, the
disjoint uinon of two lines $S^{0}\times D^{1}=D^{1}\sqcup D^{1}$.
Finally one has to add a 2-handle $D^{2}\times D^{0}$ to get the
closed manifold $T^{2}$. Every 1-handle can be covered by (at least)
two charts and finally we recover the covering by 4 charts. Both pictures
are equivalent but the handle picture has one simple advantage: it
reduces the number of fundamental pieces of a manifold and of the
transition maps. The gluing maps of the handles can be seen as a generalization
of transition maps. Then the handle picture presents only the most
important of these gluing or transition maps, omitting the trivial
transition maps.

In this paper we will present such a coordinate representation, albeit
infinite, of an exotic $\mathbb{R}^{4}$ based on the handle body
decomposition of Bizaca and Gompf. We suggest that one of the consequences
of this approach would be to suggest a positive answer for the Brans
conjecture \cite{AsselmeyerBrans2012}, that exotic smoothness serves
as an additional gravitational source as a spinor field naturally
arising from the handlebody construction. The compact case was worked
out in \cite{AsselmeyerRose2012}. This might be considered as a construction
analogous to using the metric as a physical field once Einstein thought
to look at gravity as a geometric effect. In other words, if we look
for exotic smoothness effects in physics, the appearance of the spinor
field in the periodic end construction parallels Einstein's looking
to geometry as physics and then choosing the metric for gravity.

\section{Construction of exotic $\mathbb{R}^{4}$}

Our model of space-time is the non-compact space topological $\mathbb{R}^{4}$.
The results can be easily generalized for other cases such as $S^{3}\times\mathbb{R}$.
In this section we will give some information about the construction
of exotic $\mathbb{R}^{4}$. The existence of a smooth embedding $R^{4}\to S^{4}$
of the exotic $\mathbb{R}^{4}$ into the 4-sphere splits all exotic
$\mathbb{R}^{4}$ into two classes, large (no embedding) or small.

\subsection{Preliminaries: Slice and non-slice knots}

\begin{figure}
\begin{center}\includegraphics[scale=0.5]{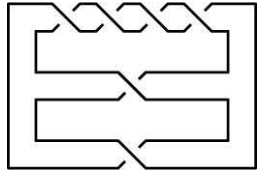}\end{center}

\caption{a slice knot: Stevedore's knot $6_{1}$\label{fig:Stevedore-knot-6_1}}
\end{figure}

At first we start with some definitions from knot theory. A (smooth)
knot $K$ is a smooth embedding $S^{1}\to S^{3}$. In the following
we assume every knot to be smooth. Secondly we exclude wilderness
of knots, i.e the knot is equivalent to a polygon in $\mathbb{R}^{3}$
or $S^{3}$ (tame knot). Furthermore, the $n$-disk is denoted by
$D^{n}$ with $\partial D^{n}=S^{n-1}$. \begin{definition} \textbf{Smoothly
Slice Knot}: A knot in $\partial D^{4}=S^{3}$ is smoothly slice if
there exists a two-disk $D^{2}$ smoothly embedded in $D^{4}$ such
that the image of $\partial D^{2}=S^{1}$ is $K$. \end{definition}
An example of a slice knot is the so-called Stevedore's Knot (in Rolfson
notation $6_{1}$, see Fig. \ref{fig:Stevedore-knot-6_1}). 

\begin{figure}
\begin{center}\includegraphics[angle=90]{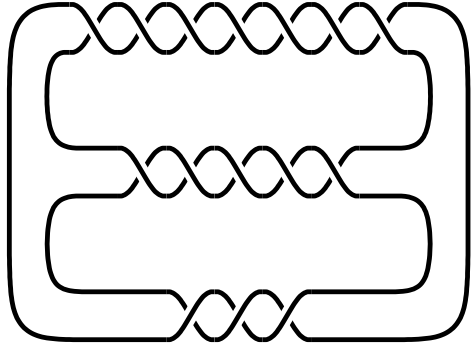}\end{center}

\caption{topological, non-smoothly slice knot: pretzel knot $(-3,5,7)$\label{fig:pretzel-knot-3-5-7}}
\end{figure}

\begin{definition} \textbf{Flat Topological Embedding}: Let $X$
be a topological manifold of dimension $n$ and $Y$ a topological
manifold of dimension $m$ where $n<m$. A topological embedding $\rho:X\to Y$
is flat if it extends to a topological embedding $\rho:X\times D^{m-n}\to Y$.

\textbf{Topologically Slice Knot}: A knot $K$ in $\partial D^{4}$
is topologically slice if there exists a two-disk $D^{2}$ flatly
topologically embedded in $D^{4}$ such that the image of $\partial D^{2}$
is $K$. \end{definition} Here we remark that the flatness condition
is essential. Any knot $K\subset S^{3}$ is the boundary of a disc
$D^{2}$ embedded in $D^{4}$, which can be seen by taking the cone
over the knot. But the vertex of the cone is a non-flat point (the
knot is crashed to a point). The difference between the smooth and
the flat topological embedding is the key for the following discussion.
This innocent looking difference seem to imply that both definitions
are equivalent. But deep results from 4-manifold topology gave a negative
answer: there are topologically slice knots which are not smoothly
slice. An example is the pretzel knot $(-3,5,7)$ (see Fig. \ref{fig:pretzel-knot-3-5-7}).

In \cite{Fre:82a}, Freedman gave a topological criteria for topological
sliceness: the Alexander polynomial $\triangle_{K}(t)$ (the best
known knot invariant, see \cite{Rol:76}) of the knot $K$ has to
be one, $\triangle_{K}(t)=1$. An example how to measure the smooth
sliceness is given by the smooth 4-genus $g_{4}(K)$ of the knot $K$,
i.e. the minimal genus of a surface $F$ smoothly embedded in $D^{4}$
with boundary $\partial F=K$ the knot. This surface $F$ is called
the \emph{Seifert surface}. Therefore, if the smooth 4-genus vanishes
$g_{4}(K)=0$ then the knot $K$ bounds a 2-disk $D^{2}$ (surface
of genus $0$) given by the smooth embedding $D^{2}\to D^{4}$ so
that the image of $\partial D^{2}\to\partial D^{4}$ is the knot $K$.

\subsection{Large exotic $\mathbb{R}^{4}$ and non-slice knots\label{sub:Large-exotic-R4}}

Large exotic $\mathbb{R}^{4}$ can be constructed by using the failure
to arbitrarily split of a compact, simple-connected 4-manifold. For
every topological 4-manifold one knows how to split this manifold
\emph{topologically} into simpler pieces using the work of Freedman
\cite{Fre:82}. But as shown by Donaldson \cite{Don:83}, some of
these 4-manifolds do not exist as smooth 4-manifolds. This contradiction
between the continuous and the smooth case produces the first examples
of exotic $\mathbb{R}^{4}$\cite{Gom:83}. Unfortunately, the construction
method is rather indirect and therefore useless for the calculation
of the path integral contribution of the exotic $\mathbb{R}^{4}$.
But as pointed out by Gompf (see \cite{Gom:85} or \cite{GomSti:97}
Exercise 9.4.23 on p. 377ff and its solution on p. 522ff), large exotic
$\mathbb{R}^{4}$ can be also constructed by using smoothly non-slice
but topologically slice knots. Especially one obtains an explicit
construction which will be used in the calculations later.

Let $K$ be a knot in $\partial D^{4}$ and $X_{K}$ the two-handlebody
obtained by attaching a two-handle to $D^{4}$ along $K$ with framing
$0$. That means: one has a two-handle $D^{2}\times D^{2}$ which
is glued to the 0-handle $D^{4}$ along its boundary using a map $f:\partial D^{2}\times D^{2}\to\partial D^{4}$
so that $f(.\,,\, x)=K\times x\subset S^{3}=\partial D^{4}$ for all
$x\in D^{2}$ (or the image $im(f)=K\times D^{2}$ is the solid knotted
torus). Let $\rho:X_{K}\to\mathbb{R}^{4}$ be a flat topological embedding
($K$ is topologically slice). For $K$ a smoothly non-slice knot,
the open 4-manifold 
\begin{equation}
R^{4}=\left(\mathbb{R}^{4}\setminus int\rho(X_{K})\right)\cup_{\partial X_{K}}X_{K}\label{eq:decomposition-large-exotic-R4}
\end{equation}
where $int\rho(X_{K})$ is the interior of $\rho(X_{K})$, is homeomorphic
but non-diffeomorphic to $\mathbb{R}^{4}$ with the standard smoothness
structure (both pieces are glued along the common boundary $\partial X_{K}$).
The proof of this fact ($R^{4}$ is exotic) is given by contradiction,
i.e. let us assume $R^{4}$ is diffeomorphic to $\mathbb{R}^{4}$.
Thus, there exists a diffeomorphism $R^{4}\to\mathbb{R}^{4}$. The
restriction of this diffeomorphism to $X_{K}$ is a smooth embedding
$X_{K}\hookrightarrow\mathbb{R}^{4}$. However, such a smooth embedding
exists if and only if $K$ is smoothly slice (see \cite{GomSti:97}).
But, by hypothesis, $K$ is not smoothly slice. Thus by contradiction,
there exists a no diffeomorphism $R^{4}\to\mathbb{R}^{4}$ and $R^{4}$
is exotic, homeomorphic but not diffeomorphic to $\mathbb{R}^{4}$.
Finally, we have to prove that $R^{4}$ is large. $X_{K}$, by construction,
is compact and a smooth submanifold of $R^{4}$. By hypothesis, $K$
is not smoothly slice and therefore $X_{K}$ can not smoothly embed
in $\mathbb{R}^{4}$. By restriction, $D^{4}\subset X_{K}$ and also
$\partial D^{4}=S^{3}$ can not smoothly embed and therefore $R^{4}$
is a large exotic $\mathbb{R}^{4}$.

\subsection{Small exotic $\mathbb{R}^{4}$ and Casson handles\label{sub:Small-exotic-R4}}

Small exotic $\mathbb{R}^{4}$'s are again the result of anomalous
smoothness in 4-dimensional topology but of a different kind than
for large exotic $\mathbb{R}^{4}$'s. In 4-manifold topology \cite{Fre:82},
a homotopy-equivalence between two compact, closed, simply-connected
4-manifolds implies a homeomorphism between them (a so-called h cobordism).
But Donaldson \cite{Don:87} provided the first smooth counterexample,
i.e. both manifolds are generally not diffeomorphic to each other.
The failure can be localized in some contractible submanifold (Akbulut
cork) so that an open neighborhood of this submanifold is a small
exotic $\mathbb{R}^{4}$. The whole procedure implies that this exotic
$\mathbb{R}^{4}$ can be embedded in the 4-sphere $S^{4}$.

The idea of the construction is simply given by the fact that every
such smooth h-cobordism between non-diffeomorphic 4-manifolds can
be written as a product cobordism except for a compact contractible
sub-h-cobordism $V$, the Akbulut cork. An open subset $U\subset V$
homeomorphic to $[0,1]\times{{\mathbb{R}}^{4}}$ is the corresponding
sub-h-cobordism between two exotic ${{\mathbb{R}}^{4}}$'s. These
exotic ${{\mathbb{R}}^{4}}$'s are called ribbon ${{\mathbb{R}}^{4}}$'s.
They have the important property of being diffeomorphic to open subsets
of the standard ${{\mathbb{R}}^{4}}$. To be more precise, consider
a pair $(X_{+},X_{-})$ of homeomorphic, smooth, closed, simply-connected
4-manifolds.

\begin{theorem}\emph{ }Let $W$ be a smooth h-cobordism between closed,
simply connected 4-manifolds $X_{-}$ and $X_{+}$. Then there is
an open subset $U\subset W$ homeomorphic to $[0,1]\times{{\mathbb{R}}^{4}}$
with a compact subset $C\subset U$ such that the pair $(W\setminus C,U\setminus C)$
is diffeomorphic to a product $[0,1]\times(X_{-}\setminus C,U\cap X_{-}\setminus C)$.
The subsets $R_{\pm}=U\cap X_{\pm}$ (homeomorphic to ${{\mathbb{R}}^{4}}$)
are diffeomorphic to open subsets of ${{\mathbb{R}}^{4}}$. If $X_{-}$
and $X_{+}$ are not diffeomorphic, then there is no smooth 4-ball
in $R_{\pm}$ containing the compact set $Y_{\pm}=C\cap R_{\pm}$,
so both $R_{\pm}$ are exotic ${{\mathbb{R}}^{4}}$'s.\emph{ }\end{theorem}

Thus, remove a certain contractible, smooth, compact 4-manifold $Y_{-}\subset X_{-}$
(called an Akbulut cork) from $X_{-}$, and re-glue it by an involution
of $\partial Y_{-}$, i.e. a diffeomorphism $\tau:\partial Y_{-}\to\partial Y_{-}$
with $\tau\circ\tau=Id$ and $\tau(p)\not=\pm p$ for all $p\in\partial Y_{-}$.
This argument was modified above so that it works for a contractible
{\em open} subset $R_{-}\subset X_{-}$ with similar properties,
such that $R_{-}$ will be an exotic ${{\mathbb{R}}^{4}}$ if $X_{+}$
is not diffeomorphic to $X_{-}$. Furthermore $R_{-}$ lies in a compact
set, i.e. a 4-sphere or $R_{-}$ is a small exotic $\mathbb{R}^{4}$.
In \cite{DeMichFreedman1992} Freedman and DeMichelis constructed
also a continuous family of small exotic $\mathbb{R}^{4}$.

Now we are ready to discuss the decomposition of a small exotic $\mathbb{R}^{4}$
by Bizaca and Gompf \cite{BizGom:96} by using special pieces, the
handles forming a handle body. Every 4-manifold can be decomposed
(seen as handle body) using standard pieces such as $D^{k}\times D^{4-k}$,
the so-called $k$-handle attached along $\partial D^{k}\times D^{4-k}$
to the boundary $S^{3}=\partial D^{4}$ of a $0-$handle $D^{0}\times D^{4}=D^{4}$.
The construction of the handle body can be divided into two parts.
The first part is known as the Akbulut cork, a contractable 4-manifold
with boundary a homology 3-sphere (a 3-manifold with the same homology
as the 3-sphere). The Akbulut cork $A_{cork}$ is given by a linking
between a 1-handle and a 2-handle of framing $0$. The second part
is the Casson handle $CH$ which will be considered now.

Let us start with the basic construction of the Casson handle $CH$.
Let $M$ be a smooth, compact, simple-connected 4-manifold and $f:D^{2}\to M$
a (codimension-2) mapping. By using diffeomorphisms of $D^{2}$ and
$M$, one can deform the mapping $f$ to get an immersion (i.e. injective
differential) generically with only double points (i.e. $\#|f^{-1}(f(x))|=2$)
as singularities \cite{GolGui:73}. But to incorporate the generic
location of the disk, one is rather interesting in the mapping of
a 2-handle $D^{2}\times D^{2}$ induced by $f\times id:D^{2}\times D^{2}\to M$
from $f$. Then every double point (or self-intersection) of $f(D^{2})$
leads to self-plumbings of the 2-handle $D^{2}\times D^{2}$. A self-plumbing
is an identification of $D_{0}^{2}\times D^{2}$ with $D_{1}^{2}\times D^{2}$
where $D_{0}^{2},D_{1}^{2}\subset D^{2}$ are disjoint sub-disks of
the first factor disk%
\footnote{In complex coordinates the plumbing may be written as $(z,w)\mapsto(w,z)$
or $(z,w)\mapsto(\bar{w},\bar{z})$ creating either a positive or
negative (respectively) double point on the disk $D^{2}\times0$ (the
core).%
}. Consider the pair $(D^{2}\times D^{2},\partial D^{2}\times D^{2})$
and produce finitely many self-plumbings away from the attaching region
$\partial D^{2}\times D^{2}$ to get a kinky handle $(k,\partial^{-}k)$
where $\partial^{-}k$ denotes the attaching region of the kinky handle.
A kinky handle $(k,\partial^{-}k)$ is a one-stage tower $(T_{1},\partial^{-}T_{1})$
and an $(n+1)$-stage tower $(T_{n+1},\partial^{-}T_{n+1})$ is an
$n$-stage tower union kinky handles $\bigcup_{\ell=1}^{n}(T_{\ell},\partial^{-}T_{\ell})$
where two towers are attached along $\partial^{-}T_{\ell}$. Let $T_{n}^{-}$
be $(\mbox{interior}T_{n})\cup\partial^{-}T_{n}$ and the Casson handle
\[
CH=\bigcup_{\ell=0}T_{\ell}^{-}
\]
is the union of towers (with direct limit topology induced from the
inclusions $T_{n}\hookrightarrow T_{n+1}$).

The main idea of the construction above is very simple: an immersed
disk (disk with self-intersections) can be deformed into an embedded
disk (disk without self-intersections) by sliding one part of the
disk along another (embedded) disk to kill the self-intersections.
Unfortunately the other disk can be immersed only. But the immersion
can be deformed to an embedding by a disk again etc. In the limit
of this process one ''shifts the self-intersections into infinity''
and obtains%
\footnote{In the proof of Freedman \cite{Fre:82}, the main complications come
from the lack of control about this process. %
} the standard open 2-handle $(D^{2}\times\mathbb{R}^{2},\partial D^{2}\times\mathbb{R}^{2})$.

A Casson handle is specified up to (orientation preserving) diffeomorphism
(of pairs) by a labeled finitely-branching tree with base-point {*},
having all edge paths infinitely extendable away from {*}. Each edge
should be given a label $+$ or $-$. Here is the construction: tree
$\to CH$. Each vertex corresponds to a kinky handle; the self-plumbing
number of that kinky handle equals the number of branches leaving
the vertex. The sign on each branch corresponds to the sign of the
associated self plumbing. The whole process generates a tree with
infinite many levels. In principle, every tree with a finite number
of branches per level realizes a corresponding Casson handle. Each
building block of a Casson handle, the ``kinky'' handle with $n$
kinks%
\footnote{The number of end-connected sums is exactly the number of self intersections
of the immersed two handle.%
}, is diffeomorphic to the $n-$times boundary-connected sum $\natural_{n}(S^{1}\times D^{3})$
(see appendix \ref{sec:Connected-and-boundary-connected}) with two
attaching regions. Technically speaking, one region is a tubular neighborhood
of band sums of Whitehead links connected with the previous block.
The other region is a disjoint union of the standard open subsets
$S^{1}\times D^{2}$ in $\#_{n}S^{1}\times S^{2}=\partial(\natural_{n}S^{1}\times D^{3})$
(this is connected with the next block).

\subsection{The Einstein-Hilbert action}

In this section we will discuss the Einstein-Hilbert action functional
\begin{equation}
S_{EH}(M)=\intop_{M}R\sqrt{g}\: d^{4}x\label{eq:EH-action}
\end{equation}
of the 4-manifold $M$ and fix the Ricci-flat metric $g$ as solution
of the vacuum field equations of the exotic 4-manifold. The main part
of our argumentation is additional contribution to the action functional
coming from exotic smoothness.

In case of the large exotic $\mathbb{R}^{4}$, we consider the decomposition
\begin{eqnarray}
R^{4} & = & \left(\mathbb{R}^{4}\setminus int\rho(X_{K})\right)\cup_{\partial X_{K}}X_{K}\label{eq:relation-exotic}
\end{eqnarray}
where $R^{4}$ is the large exotic $\mathbb{R}^{4}$. For the parts
of the decomposition we obtain the action functionals 
\begin{eqnarray*}
S_{EH}(\mathbb{R}^{4}\setminus int\rho(X_{K})) & = & \intop_{\mathbb{R}^{4}\setminus int\rho(X_{K})}R\sqrt{g}\, d^{4}x+\intop_{\partial X_{K}}H\sqrt{h}d^{3}x\\
S_{EH}(X_{K}) & = & \intop_{X_{K}}R\sqrt{g}d^{4}x-\intop_{\partial X_{K}}H\sqrt{h}d^{3}x
\end{eqnarray*}
including the contribution of the boundary $\partial X_{K}$ with
respect to different orientations and $H$ is the trace of the second
fundamental form (mean curvature) of the boundary in the metric $g$.

As explained above, a small exotic $\mathbb{R}^{4}$ can be decomposed
into a compact subset $A_{cork}$ (Akbulut cork) and a Casson handle
(see \cite{BizGom:96}). Especially this exotic $\mathbb{R}^{4}$
depends strongly on the Casson handle, i.e. non-diffeomorphic Casson
handles lead to non-diffeomorphic $\mathbb{R}^{4}$'s. Thus we have
to understand the analytical properties of a Casson handle. In \cite{Kato2004},
the analytical properties of the Casson handle were discussed. The
main idea is the usage of the theory of end-periodic manifolds, i.e.
an infinite periodic structure generated by $W$ glued along a compact
set $A_{cork}$ to get for the interior 
\[
\mathbb{R}_{\theta}^{4}=int\left(A_{cork}\cup_{N}W\cup_{N}W\cup_{N}\cdots\right)
\]
the end-periodic manifold. The definition of an end-periodic manifold
is very formal (see \cite{Tau:87}) and we omit it here. All Casson
handles generated by a balanced tree have the structure of end-periodic
manifolds as shown in \cite{Kato2004}. By using the theory of Taubes
\cite{Tau:87} one can construct a metric on $\cdots\cup_{N}W\cup_{N}W\cup_{N}\cdots$
by using the metric on $W$. Then a metric $g$ in $\mathbb{R}_{\theta}^{4}$
transforms to a periodic function $\hat{g}$ on the infinite periodic
manifold 
\[
\tilde{Y}=\cdots\cup_{N}W_{-1}\cup_{N}W_{0}\cup_{N}W_{1}\cup_{N}\cdots
\]
where $W_{i}$ is the building block $W$ at the $i$th place. Then
the action of $\mathbb{R}_{\theta}^{4}$ can be divided into many
parts 
\begin{eqnarray*}
S_{EH}(A_{cork}) & = & \intop_{A_{cork}}R\sqrt{g}d^{4}x-\intop_{\partial A_{cork}}H\,\sqrt{h}d^{3}x\\
S_{EH}(W_{i}) & = & \intop_{W_{i}}R\sqrt{g}d^{4}x+\intop_{N}H\,\sqrt{h}d^{3}x
\end{eqnarray*}
again including the boundaries $N=\partial W$ and $\partial A_{cork}$.
In any case we can reduce the problem to the discussion of the action
\begin{equation}
S_{EH}(\Sigma)=\intop_{\Sigma}H\,\sqrt{h}\, d^{3}x\label{eq:action fermi}
\end{equation}
along the boundary $\Sigma$ (a 3-manifold). It is a surprise that
this integral agrees with the Dirac action of a spinor describing
the (immersed) boundary, see below.

\section{Immersed surfaces and the Dirac action\label{sec:Dirac-action}}

In the following we will show that the action (\ref{eq:action fermi})
is completely determined by the knotted torus $\partial N(K)=K\times S^{1}$
and its mean curvature $H_{\partial N(K)}$. This knotted torus is
an immersion of a torus $S^{1}\times S^{1}$ into $\mathbb{R}^{3}$.
The well-known \emph{Weierstrass representation} can be used to describe
this immersion. As proved in \cite{SpinorRep1996,Friedrich1998} there
is an equivalent representation via spinors. This so-called \emph{Spin
representation} of a surface gives back an expression for $H_{\partial N(K)}$
and the Dirac equation as geometric condition on the immersion of
the surface. As we will show below, the term (\ref{eq:action fermi})
can be interpreted as Dirac action of a spinor field.

\subsection{From 3-manifolds to immersed surfaces\label{sub:From-3-manifolds-to-immersed-surfaces}}

\begin{figure}
\includegraphics[scale=0.3]{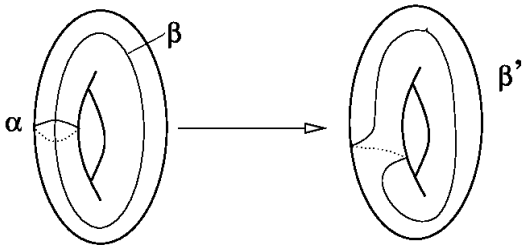}

\caption{Dehn twist\label{fig:Dehn-twist}}
\end{figure}

The action (\ref{eq:action fermi}) depends on the 3-manifold $\Sigma$
as the boundary of an appropriated 4-manifold $M$. Then the embedding
of this boundary depends on the 3-manifold $\Sigma$ which we have
to describe first. The relation between the 4-manifold and the boundary
(a 3-manifold) is very close. In particular, the 4-manifolds in this
paper can be obtained by adding 2-handles (glued to the 0-handle by
using knots). Then one can construct the 3-manifold by similar methods.
The core of this method is the following result: Let $\Sigma$ be
an arbitrary 3-manifold and $S^{3}$ the 3-sphere. Cut out a solid
torus $T=S^{1}\times D^{2}$ from both manifolds then $\Sigma\setminus T$
and $S^{3}\setminus T$ are homeomorphic (and also diffeomorphic).
So, every 3-manifold can be generated by a procedure (called surgery):
cutting out solid tori from the 3-sphere $S^{3}$ and then pasting
them back in, but along different homeomorphisms of their boundaries.
Then the homeomorphisms of the boundaries, the usual torus $T^{2}=S^{1}\times S^{1}$,
determine the 3-manifold completely. Homeomorphisms of the torus $T^{2}$
are well understood using Dehn twists. In a Dehn twist, one cut the
torus to obtain a cylinder and past both ends together after a full
twist of one end (see Fig. \ref{fig:Dehn-twist}). 

Equally one can also do a twist along the other curve $\alpha$. For
a coordinate description of this procedure one considers the torus
as a product of two circles $S^{1}\times S^{1}$ (denoted by $\alpha,\beta$
in the Fig.). Let $(\theta,\phi)$ be two angle coordinates (range
$[0,2\pi),$ for each $S^{1}$ factor. If $u(\alpha)$ is a smooth
function equal to one near $\pi$ but zero elsewhere, we represent
the $(p,q)$ twist by 
\[
(\theta,\phi)\to(\theta+p2\pi u(\phi),\phi+q2\pi u(\theta)).
\]
Or, we can define the $(p,q)\in\mathbb{Z}^{2}$ twisted torus as identification
space resulting from identifying $(x,y)\sim(mpx,nqy)$ for any $(m,n)\in\mathbb{Z}^{2}$
and $(x,y)\in\mathbb{R}^{2}$. By a Dehn twist, one obtains a knotting
of the torus. But more importantly we obtain a surface inside of the
3-manifold (unique up to diffeomorphisms) so that the embedding of
the 3-manifold can be described by the embedding of this surface.
Therefore, two 3-manifolds agree on the complement of some disjoint
solid tori, i.e. we have for a 3-manifold $\Sigma=(S^{3}\setminus(D^{2}\times S^{1}))\cup(K\times D^{2})$
for some knot $K$. The first contribution $S^{3}\setminus(D^{2}\times S^{1})=S^{1}\times D^{2}$
can be chosen using a constant embedding, i.e. we have for the integral
(\ref{eq:action fermi}) 
\begin{equation}
S_{EH}(\Sigma)=\intop_{\Sigma}H\,\sqrt{h}\, d^{3}x=\intop_{K\times D^{2}}H\,\sqrt{h}\, d^{3}x+const.+boundary\label{eq:action fermi new}
\end{equation}
where the contribution of $N(K)=K\times D^{2}$ reflects the dependence
on the topology of the 3-manifold $\Sigma$. The integral (\ref{eq:action fermi})
for two different 3-manifolds differs exactly by this expression.
From the topological point of view, we can write alternatively 
\[
K\times D^{2}=K\times D^{1}\times D^{1}=K\times[0,1]^{2}\,.
\]
Finally we end up with the action 
\begin{equation}
S_{EH}(K\times D^{1}\times D^{1})=\intop_{K\times D^{1}\times D^{1}}H\,\sqrt{h}\, d^{3}x\quad.\label{eq:action-fermi-rewrite}
\end{equation}
Obviously, the complexity of the embedding is given by the knot $K$
or by a plane like $K\times D^{1}$. Without loss of generality, we
choose a product metric and consider the mean curvature $H_{K}$ for
the embedding $K\times D^{1}$ to state 
\begin{equation}
S_{EH}(K\times D^{1}\times D^{1})=\intop_{D^{1}}d\theta\intop_{K\times D^{1}}H_{K}\,\sqrt{h}\, d^{2}x\quad.\label{eq:action-fermi-final}
\end{equation}

\subsection{Weierstrass and spin representation of immersed submanifolds\label{subsubsec:Spin-representation-disk}}

In this subsection we will describe the theory of immersions using
spinors. The theory will be presented stepwise. We start with a toy
model of an immersion of a surface into the 3-dimensional Euclidean
space. Then we discuss how this map can be extended to an immersion
of a 3-manifold into a 4-manifold.

Let $f:M^{2}\to\mathbb{R}^{3}$ be a smooth map of a Riemannian surface
with injective differential $df:TM^{2}\to T\mathbb{R}^{3}$, i.e.
an immersion. In the \emph{Weierstrass representation} one expresses
a \emph{conformal minimal} immersion $f$ in terms of a holomorphic
function $g\in\Lambda^{0}$ and a holomorphic 1-form $\mu\in\Lambda^{1,0}$
as the integral 
\[
f=Re\left(\int(1-g^{2},i(1+g^{2}),2g)\mu\right)\ .
\]
An immersion of $M^{2}$ is conformal if the induced metric $g$ on
$M^{2}$ has components 
\[
g_{zz}=0=g_{\bar{z}\bar{z}}\,,\: g_{z\bar{z}}\not=0
\]
and it is minimal if the surface has minimal volume. Now we consider
a spinor bundle $S$ on $M^{2}$ (i.e. $TM^{2}=S\otimes S$ as complex
line bundles) and with the splitting 
\[
S=S^{+}\oplus S^{-}=\Lambda^{0}\oplus\Lambda^{1,0}
\]
Therefore the pair $(g,\mu)$ can be considered as spinor field $\varphi$
on $M^{2}$. Then the Cauchy-Riemann equation for $g$ and $\mu$
is equivalent to the Dirac equation $D\varphi=0$. The generalization
from a conformal minimal immersion to a conformal immersion was done
by many authors (see the references in \cite{Friedrich1998}) to show
that the spinor $\varphi$ now fulfills the Dirac equation 
\begin{equation}
D\varphi=H\varphi\label{eq:conformal-immersion-Dirac}
\end{equation}
where $H$ is the mean curvature (i.e. the trace of the second fundamental
form). The minimal case is equivalent to the vanishing mean curvature
$H=0$ recovering the equation above. Friedrich \cite{Friedrich1998}
uncovered the relation between a spinor $\Phi$ on $\mathbb{R}^{3}$
and the spinor $\varphi=\Phi|_{M^{2}}$: if the spinor $\Phi$ fulfills
the Dirac equation $D\Phi=0$ then the restriction $\varphi=\Phi|_{M^{2}}$
fulfills equation (\ref{eq:conformal-immersion-Dirac}) and $|\varphi|^{2}=const$.
Therefore we obtain 
\begin{equation}
H=\bar{\varphi}D\varphi\label{eq:mean-curvature-surface}
\end{equation}
with $|\varphi|^{2}=1$.

After this exercise we are ready to consider the integral (\ref{eq:action-fermi-final}).
Here we have an immersion of $I:S^{1}\times D^{1}\to\mathbb{R}^{3}$
with image the thicken knot $im(I)=T(K)=K\times D^{1}$. This immersion
$I$ can be defined by a spinor $\varphi$ on $S^{1}\times D^{1}$
fulfilling the Dirac equation 
\begin{equation}
D\varphi=H\varphi\label{eq:2D-Dirac}
\end{equation}
with $|\varphi|^{2}=1$ (or an arbitrary constant) (see Theorem 1
of \cite{Friedrich1998}). As discussed above a spinor bundle over
a surface splits into two sub-bundles $S=S^{+}\oplus S^{-}$ with
the corresponding splitting of the spinor $\varphi$ in components
\[
\varphi=\left(\begin{array}{c}
\varphi^{+}\\
\varphi^{-}
\end{array}\right)
\]
and we have the Dirac equation 
\[
D\varphi=\left(\begin{array}{cc}
0 & \partial_{z}\\
\partial_{\bar{z}} & 0
\end{array}\right)\left(\begin{array}{c}
\varphi^{+}\\
\varphi^{-}
\end{array}\right)=H\left(\begin{array}{c}
\varphi^{+}\\
\varphi^{-}
\end{array}\right)
\]
with respect to the coordinates $(z,\bar{z})$ on $S^{1}\times D^{1}$.

In dimension 3, the spinor bundle has the same fiber dimension as
the spinor bundle $S$ (but without a splitting $S=S^{+}\oplus S^{-}$into
two sub-bundles). Now we define the extended spinor $\phi$ over the
solid torus $T^{3}=S^{1}\times D^{1}\times D^{1}=S^{1}\times D^{2}$
via the restriction $\phi|_{S^{1}\times D^{1}}=\varphi$. The spinor
$\phi$ is constant along the normal vector $\partial_{N}\phi=0$
fulfilling the 3-dimensional Dirac equation 
\begin{equation}
D^{3D}\phi=\left(\begin{array}{cc}
\partial_{N} & \partial_{z}\\
\partial_{\bar{z}} & -\partial_{N}
\end{array}\right)\phi=H\phi\label{eq:Dirac-equation-3D}
\end{equation}
induced from the Dirac equation (\ref{eq:2D-Dirac}) via restriction
and where $|\phi|^{2}=const.$ Especially one obtains for the mean
curvature of the knotted solid torus $K\times D^{2}$ (up to a constant
from $|\phi|^{2}$) 
\begin{equation}
H=\bar{\phi}D^{3D}\phi\,.\label{eq:mean-curvature-3D}
\end{equation}

\subsection{Deformation of the Immersion and the spectrum of the Dirac operator}

Now we will discuss the change of the immersion by a diffeomorphism.
But first, we will remark that (\ref{eq:2D-Dirac}) and (\ref{eq:Dirac-equation-3D})
are eigenvalue equations. The eigenvectors correspond to immersions
where the eigenvalue is the mean curvature of this immersion. Then
any other immersion corresponds to a linear combination of eigenvectors.
The mean curvature of this immersion is also a linear combination
of the eigenvalues. In particular, there is also the eigenvector to
the eigenvalue $0$, called the minimal immersion. Thus, we obtain
a quantized (mean) curvature as eigenvalues of a Dirac operator. This
approach has some similarities with the spectral triple in noncommutative
geometry \cite{Con:95}. But in contrast to noncommutative geometry,
we start with the simple model to use an exotic smoothness structure.
Why did we obtain a similar result? There are many hints that an exotic
$\mathbb{R}^{4}$ is a noncommutative space in the sense of Connes.
We partly worked out this theory using wild embeddings \cite{AsselmeyerKrol2013}.

Now we will discuss the deformation of a immersion using a diffeomorphism.
Let $I:\Sigma\hookrightarrow M$ be an immersion of $\Sigma$ (3-manifold)
into $M$ (4-manifold). A deformation of an immersion $I':\Sigma'\hookrightarrow M'$
are diffeomorphisms $f:M\to M'$ and $g:\Sigma\to\Sigma'$ of $M$
and $\Sigma$, respectively, so that 
\[
I\circ f=g\circ I'\,.
\]
One of the diffeomorphism (say $f$) can be absorbed into the definition
of the immersion and we are left with one diffeomorphism $g\in Diff(\Sigma)$
to define the deformation of the immersion $I$. But as stated above,
the immersion is directly given by an integral over the spinor $\phi$
on $\Sigma$ fulfilling the Dirac equation (\ref{eq:Dirac-equation-3D}).
Therefore we have to discuss the action of the diffeomorphism group
$Diff(\Sigma)$ on the Hilbert space of $L^{2}-$spinors fulfilling
the Dirac equation. This case was considered in the literature \cite{SpinorsDiffeom2013}.
The spinor space $S_{g,\sigma}(\Sigma)$ on $\Sigma$ depends on two
ingredients: a (Riemannian) metric $g$ and a spin structure $\sigma$
(labeled by the number of elements in $H^{1}(\Sigma,\mathbb{Z}_{2})$).
Let us consider the group of orientation-preserving diffeomorphism
$Diff^{+}(\Sigma)$ acting on $g$ (by pullback $f^{*}g$) and on
$\sigma$ (by a suitable defined pullback $f^{*}\sigma$). The Hilbert
space of $L^{2}-$spinors of $S_{g,\sigma}(\Sigma)$ is denoted by
$H_{g,\sigma}$. Then according to \cite{SpinorsDiffeom2013}, any
$f\in Diff^{+}(\Sigma)$ leads in exactly two ways to a unitary operator
$U$ from $H_{g,\sigma}$ to $H_{f^{*}g,f^{*}\sigma}$. The (canonically)
defined Dirac operator is equivariant with respect to the action of
$U$ and the spectrum is invariant under (orientation-preserving)
diffeomorphisms. But by the discussion above, we also do not change
the immersion by a diffeomorphism. So, our whole approach is independent
on a concrete coordinate system.

\subsection{The Dirac action in 3 dimensions and the 4-dimensional Dirac equation}

By using the relation (\ref{eq:mean-curvature-3D}) above we obtain
for the integral (\ref{eq:action fermi}) 
\begin{equation}
\intop_{K\times D^{2}}H_{K}\sqrt{h}d\theta d^{2}x=\intop_{K\times D^{2}}\bar{\phi}D^{3D}\phi\:\sqrt{h}\, d\theta d^{2}x\label{eq:3D-action-fermion}
\end{equation}
i.e. the Dirac action on the knotted solid torus $K\times D^{2}=T^{3}(K)$.
But that is not the expected result, we obtain only a 3-dimensional
Dirac action leaving us with the question to extend the action to
four dimensions.

Let $\iota:T^{3}\hookrightarrow M$ be an immersion of the solid torus
$\Sigma=T^{3}$ into the 4-manifold $M$ with the normal vector $\vec{N}$.
At this stage one can consider an arbitrary 3-manifold $\Sigma$ instead
of the 3-torus. The spin bundle $S_{M}$ of the 4-manifold splits
into two sub-bundles $S_{M}^{\pm}$ where one subbundle, say $S_{M}^{+},$
can be related to the spin bundle $S_{\Sigma}$ of the 3-manifold.
Then the spin bundles are related by $S_{\Sigma}=\iota^{*}S_{M}^{+}$
with the same relation $\phi=\iota_{*}\Phi$ for the spinors ($\phi\in\Gamma(S_{\Sigma})$
and $\Phi\in\Gamma(S_{M}^{+})$). Let $\nabla_{X}^{M},\nabla_{X}^{\Sigma}$
be the covariant derivatives in the spin bundles along a vector field
$X$ as section of the bundle $T\Sigma$. Then we have the formula
\begin{equation}
\nabla_{X}^{M}(\Phi)=\nabla_{X}^{\Sigma}\phi-\frac{1}{2}(\nabla_{X}\vec{N})\cdot\vec{N}\cdot\phi\label{eq:covariant-derivative-immersion}
\end{equation}
with the obvious embedding $\phi\mapsto\left(\begin{array}{c}
\phi\\
0
\end{array}\right)=\Phi$ of the spinor spaces. The expression $\nabla_{X}\vec{N}$ is the
second fundamental form of the immersion where the trace $tr(\nabla_{X}\vec{N})=2H$
is related to the mean curvature $H$. Then from (\ref{eq:covariant-derivative-immersion})
one obtains a similar relation between the corresponding Dirac operators
\begin{equation}
D^{M}\Phi=D^{3D}\phi-H\phi\label{eq:relation-Dirac-3D-4D}
\end{equation}
with the Dirac operator $D^{3D}$ defined via (\ref{eq:Dirac-equation-3D}).
Together with equation (\ref{eq:Dirac-equation-3D}) we obtain 
\begin{equation}
D^{M}\Phi=0\label{eq:Dirac-equation-4D}
\end{equation}
i.e. $\Phi$ is a parallel spinor.

\subsection{The extension to the 4-dimensional Dirac action}

Above we obtained a relation (\ref{eq:relation-Dirac-3D-4D}) between
a 3-dimensional spinor $\phi$ on the 3-manifold $\Sigma=D^{2}\times S^{1}$
fulfilling a Dirac equation $D^{\Sigma}\phi=H\phi$ (determined by
the immersion $\Sigma\to M$ into a 4-manifold $M$) and a 4-dimensional
spinor $\Phi$ on a 4-manifold $M$ with fixed chirality ($\in\Gamma(S_{M}^{+})$
or $\in\Gamma(S_{M}^{-})$) fulfilling the Dirac equation $D^{M}\Phi=0$.
At first we consider the variation 
\begin{equation}
\delta\intop_{K\times D^{2}}\bar{\phi}D^{3D}\phi\:\sqrt{g}\, d\theta d^{2}x=0\label{eq:3D-variation}
\end{equation}
of the 3-dimensional action leading to the Dirac equations 
\begin{equation}
D^{3D}\phi=0\quad D^{3D}\bar{\phi}=0\label{eq:3D-Dirac-equation}
\end{equation}
or to 
\[
H=0\,,
\]
a characterization of the immersion $K\times D^{2}$ of the solid
torus $D^{2}\times S^{1}$ with minimal mean curvature. This variation
can be understood as a variation of the (conformal) immersion. In
contrast, the extension of the spinor $\phi$ (as solution of (\ref{eq:3D-Dirac-equation}))
to the 4-dimensional spinor $\Phi$ by using the embedding 
\begin{equation}
\Phi=\left(\begin{array}{c}
\phi\\
0
\end{array}\right)\label{eq:embedding-spinor-3D-4D}
\end{equation}
can be only seen as immersion, if (and only if) the 4-dimensional
Dirac equation 
\begin{equation}
D^{M}\Phi=0\label{eq:4D-Dirac-equation}
\end{equation}
on $M$ is fulfilled (using relation (\ref{eq:relation-Dirac-3D-4D})).
This Dirac equation is obtained by varying the action 
\begin{equation}
\delta\intop_{M}\bar{\Phi}D^{M}\Phi\sqrt{g}\: d^{4}x=0\label{eq:4D-variation}
\end{equation}
Importantly, this variation has a different interpretation in contrast
to varying the 3-dimensional action. Both variations look very similar.
But in (\ref{eq:4D-variation}) we vary over smooth maps $\Sigma=D^{2}\times S^{1}\to M$
which are not conformal immersions (i.e. represented by spinors $\Phi$
with $D^{M}\Phi\not=0$). Only the choice of the extremal action selects
the conformal immersion among other smooth maps. Especially the spinor
$\Phi$ (as solution of the 4-dimensional Dirac equation) is localized
at the immersed 3-manifold $\Sigma$ (with respect to the embedding
(\ref{eq:embedding-spinor-3D-4D})). The 3-manifold $\Sigma$ moves
along the normal vector (see the relation (\ref{eq:covariant-derivative-immersion})
between the covariant derivatives representing a parallel transport).

\subsection{Matter as knot complements}

In the previous subsections we presented a formalism to describe the
immersion of a solid torus $D^{2}\times S^{1}$ with a knotted solid
torus $D^{2}\times K$ as image. Now we will go back to our original
view (see subsection \ref{sub:From-3-manifolds-to-immersed-surfaces}).
There we considered the 3-manifold $\Sigma=(S^{3}\setminus(D^{2}\times S^{1}))\cup(K\times D^{2})$
which is equally given by $\Sigma=(S^{3}\setminus(D^{2}\times K))\cup(D^{2}\times S^{1})$.
Then the spinor $\phi$ on $D^{2}\times K$ is related to the spinor
$\phi'$ on $S^{3}\setminus(D^{2}\times K)$ by a constant, which
is the normalization of the spinor $\Phi$ on $\Sigma$ with $\Phi|_{D^{2}\times K}=\phi$.
But then the spinors $\phi$ and $\phi'$ fulfill the same dynamics,
the Dirac equation. But what does it mean? From the view point of
quantum mechanics, the spinor $\phi$ as immersion of $D^{2}\times S^{1}$
is non-zero on the space of possible positions. If we make the obvious
assumtion that the complement of this space $D^{2}\times S^{1}$ is
the particle (represented by the spinor) then the particle must be
the complement $S^{3}\setminus(D^{2}\times K)$ of the knotted solid
torus. This space is also called the knot complement. A knot complement
is a compact 3-manifold with boundary a torus $T^{2}$. After the
extension to the 4-manifold $M$, the spinor $\Phi$ represents the
dynamics of the knot complement in the 4-manifold. Finally we state:\\
 \emph{Matter is represented by complements $S^{3}\setminus(D^{2}\times K)$
of knots $K$ with a dynamics determined by the Dirac equation (\ref{eq:4D-Dirac-equation}).}\\
 Currently this statement is not a large restriction. There are infinitely
many knots and we do not know which knot represents the electron or
neutrino. But for knot complements, there is a simple division into
two classes: knots with a knot complement admitting a homogenuous,
hyperbolic metric (a metric of constant negative curvature in every
direction) and knots not admitting such a metric. In \cite{AsselmeyerRose2012},
we discussed the non-hyperbolic case and showed that the corresponding
3-manifolds are representimg the interaction. Therefore we are left
with hyperbolic knot complements. In the next section we will show
that these knot complements have the right properties to describe
fermions.

\section{The physical interpretation}

In this section we will discuss the physical interpretation of the
mathematical results above including the limits of this approach.
In particular we will prove the conjecture that hyperbolic knot complements,
i.e. 3-manifolds $S^{3}\setminus\left(D^{2}\times K\right)$ admitting
a homogenuous, hyperbolic metric, representing the fermions. We used
the spinor representation to express the immersion of the submanifold.
Here we will further clarify the following questions: Does the submanifold
(the knot complement) has the properties of a spinor fulfilling the
Dirac equation? Has it also the properties of matter like non-contractability
(state equation $p=0$)? From a physical point of view, we have to
check that the submanifold (=knot complement) has 
\begin{enumerate}
\item spin $\frac{1}{2}$ (with an appropriated definition), 
\item the Dirac equationas equation of motion and 
\item the state equation $p=0$ (non-contractable matter) in the cosmological
context. 
\end{enumerate}
\textbf{ad 1.} We start with the spin. Our definition is inspired
by the work of Friedman and Sorkin \cite{FriedmanSorkin1980}, for
the details we refer to the Appendix \ref{sec:Spin-from-space}. Now
we will looking for a rotation $R(\theta)$ (rotation w.r.t. an angle
$\theta$) which acts on the 4-dimensional spinor $\Phi$. Because
of the embedding (\ref{eq:embedding-spinor-3D-4D}), it is enough
to consider the action on the 3-dimensional spinor $\phi$. Then a
rotation as element of $SO(3)$ must be represented by a diffeomorphism,
i.e. we have the representation $R:SO(3)\to Diff(\Sigma)$ where $R(\theta)$
is a one-parameter subgroup of diffeomorphisms. We call $\phi$ a
spinor if 
\[
\phi\circ R(2\pi)^{*}=-\phi\qquad\mbox{or}\qquad R(2\pi)=-1
\]
in the notation of Appendix \ref{sec:Spin-from-space}. From the topological
point of view, this rotation is located in the component of the diffeomorphism
group which is not connected to the identity. The existence of these
rotations is connecetd to the complexity of the 3-manifold. As shown
by Hendriks \cite{Hendriks1977}, these rotations do \textbf{not}
exist in sums of 3-manifolds containing 
\begin{itemize}
\item $\mathbb{R}P^{2}\times S^{1}$ with the Klein bottle $\mathbb{R}P^{2}$ 
\item $S^{2}$ fiber bundle over $S^{1}$ and 
\item for 3-manifolds with finite fundamental group having a cyclic 2-Sylow
subgroup%
\footnote{A 2-Sylow subgroup of a finite group (here the fundamental group)
is a subgroup whose order is a power of $2$ (possibly $2^{0}$) and
which is properly contained in no larger Sylow subgroup. We note that
all 2-Sylow subgroups of a given gropu are isomorphic.%
}. 
\end{itemize}
In case of hyperbolic 3-manifolds (the knot complements) one has an
infinite fundamental group and therefore it has spin $\frac{1}{2}$.\\
 \textbf{ad 2.} This part was already shown. Using the variation (\ref{eq:4D-variation})
we obtain the 4-dimensional Dirac equation (\ref{eq:4D-Dirac-equation})
in case of an immersion. Then the spinor is directly interpretable
as the immersion, see subsection \ref{subsubsec:Spin-representation-disk}.
\\
 \textbf{ad 3.} In cosmology, one has to introduce a state equation
\[
p=w\cdot\rho
\]
between the pressure and the energy density. Matter as formed by fermions
is characterized by the state equation $p=0$ or $w=0$. Equivalently,
matter is incompressible and the energy density $\rho\sim a^{-3}$
scales like the inverse volume of the 3-space w.r.t. scaling factor
$a$. The hyperbolic 3-manifold $H$, i.e. the complement of the hyperbolic
knot, has a torus boundary $T^{2}=\partial H$, i.e. $H$ admits a
hyperbolic structure in the interior only. It should also have the
property of incompressibility. But what does it mean? As a model we
consider the following 3-manifold 
\[
N=H\cup_{T^{2}}G
\]
where the two manifolds $H$ and $G$ have a common boundary, the
torus. $H$ represents the matter (by our assumption) and $G$ is
the surrounding space, i.e. we take $N$ as a model for the cosmos.
Furthermore we assume that $N$ scales w.r.t. the scaling factor $a$,
i.e. $vol(N)\sim a^{3}$. The energy density is the total energy $E_{H}$
of the matter per volume or 
\[
\rho=\frac{E_{H}}{vol(N)}\quad.
\]
The total energy $E_{H}$is related to the scalar curvature, see appendix
\ref{sec:Scalar-curvature-and-energy-density}. Using (\ref{eq:total-energy-constant-curvature}),
we obtain for the total energy of the hyperbolic 3-manifold $H$ the
total energy $E_{H}$ with 
\[
E_{H}=vol(H)\cdot\left(\frac{1}{\kappa}R_{H}+\rho_{c}\right)\quad.
\]
Therefore we will get the scaling law $\rho\sim a^{-3}$ only for
$E_{H}\sim a^{0}$ by using $vol(N)\sim a^{3}$. It is an amazing
fact that the properties of hyperbolic 3-manifolds agree with this
demand. One property of hyperbolic 3-manifolds is central: 
{\it Mostow rigidity}. As shown by Mostow \cite{Mos:68}, every hyperbolic $n-$manifold
$n>2$ of finite volume has the property: \emph{Every diffeomorphism
(especially every conformal transformation) of a hyperbolic $n-$manifold
with finite volume is induced by an isometry.} Therefore one cannot
scale a finite-volume, hyperbolic 3-manifold. Then the volume $vol(\:)$
and the curvature are topological invariants. But then $E_{H}$ is
also a topological invariant with the scaling behaviour $E_{H}\sim a^{0}$
of a topological invariant. Finally we obtain the scaling of matter
in cosmology to be $a^{-3}$ or $w=0$.\\
 Finally: \emph{Fermions are represented by hyperbolic knot complements.}

\section{The Brans conjecture: generating sources of gravity}

We only do direct geometric observations within some local, human-scaled
coordinate patch, including, of course, interpolations of signals
received from sources outside this patch. From this, we usually assume
that spacetime has the simplest global smoothness structure. Suppose
it does not, so that spacetime is exotically smooth. For example,
suppose we observe only a single mass outside our local region and
it looks like a black hole. Normally, we assume we can extrapolate
data arriving in our standard coordinate patch on earth all the way
back to the vicinity of the black hole. We ask: ''what if the smoothness
structure does not allow this?''

This question is at the core of the Brans conjecture. Exotic spacetimes
like the exotic $\mathbb{R}^{4}$ have the property that there is
no foliation like $\mathbb{R}^{3}\times\mathbb{R}$ otherwise the
spacetime has a standard smoothness structure. But all other foliations
break the strong causality, i.e. there is no unique geodesics going
in the future or past (see the discussion in \cite{AsselmeyerRose2012}).
In this paper we will go a step further and will interpret the deviation
of the smoothness structure from the standard smoothness structure
as sources of gravity. In particular we will use the theory above
to identify the sources as fermions.

\subsection{Large exotic $\mathbb{R}^{4}$}

At first we will discuss the case of a large exotic $\mathbb{R}^{4}$
as described in subsection \ref{sub:Large-exotic-R4}. Starting point
for the construction is a topologically slice but smoothly non-slice
knot $K$ (like the pretzel knot $(-3,5,7)$ in Fig. \ref{fig:pretzel-knot-3-5-7})
in $D^{4}$. Let $X_{K}$ be the two-handlebody obtained by attaching
a two-handle to $D^{4}$ along $K$ with framing $0$. Then the open
4-manifold 
\begin{equation}
R^{4}=\left(\mathbb{R}^{4}\setminus int\rho(X_{K})\right)\cup_{\partial X_{K}}X_{K}\label{eq:decomposition-large-exotic-R4-2}
\end{equation}
where $int\rho(X_{K})$ is the interior of $\rho(X_{K})$, is homeomorphic
but non-diffeomorphic to $\mathbb{R}^{4}$ with the standard smoothness
structure (both pieces are glued along the common boundary $\partial X_{K}$).
The boundary $\partial X_{K}$ can be constructed by a $0-$framed
surgery along $K$, i.e. $\partial X_{K}=\left(S^{3}\setminus\left(K\times D^{2}\right)\right)\cup_{T^{2}}D^{2}\times S^{1}$
glued along the torus respecting the framing. For the Einstein-Hilbert
action we obtain 
\begin{equation}
S_{EH}(R^{4})=\intop_{\mathbb{R}^{4}\setminus int\rho(X_{K})}R\sqrt{g}d^{4}x+\intop_{X_{K}}R\sqrt{g}d^{4}x+\intop_{\partial X_{K}}H\sqrt{h}\, d^{3}x\label{eq:EH-action-large-exotic-R4}
\end{equation}
where $H$ is the mean curvature (trace of the second fundamental
form) w.r.t. the metric $h=g|_{\partial X_{K}}$. One word about the
boundary term. Usually one obtains two boundary terms but with a different
sign. The cancellation of these terms uses implicitly the fact that
the boundary (the 3-manifold) and orientation-reversing boundary are
related by an orientation-reversing diffeomorphism so that both boundary
terms cancel. But for most 3-manifolds among them the hyperbolic 3-manifolds
it fails, i.e. there is no orientation-reversing diffeomorphism and
the two boundary contributions are different. The boundary $\partial X_{K}$
(for the pretzel knot) is also a hyperbolic 3-manifold with no orientation-reversing
diffeomorphism. Therfore we obtain a contribution from the boundary
in the action (\ref{eq:EH-action-large-exotic-R4}). By the formalism
above, we are able to construct the Dirac action on $\partial X_{K}$
\[
\intop_{\partial X_{K}}H\sqrt{h}\, d^{3}x=\intop_{\partial X_{K}}\overline{\phi}D^{3D}\phi\sqrt{h}\, d^{3}x
\]
and extend them 
\[
S_{EH}(R^{4})=\intop_{\mathbb{R}^{4}\setminus int\rho(X_{K})}R\sqrt{g}d^{4}x+\intop_{X_{K}}R\sqrt{g}d^{4}x+\intop_{R^{4}}\overline{\Phi}D\Phi\sqrt{g}d^{4}x
\]
to the whole 4-manifold (but at least to $\partial X_{K}\times[0,1]$).
Then we can simplify the action to 
\[
S_{EH}(R^{4})=\intop_{R^{4}}R\sqrt{g}d^{4}x+\intop_{R^{4}}\overline{\Phi}D\Phi\sqrt{g}d^{4}x
\]
where the spinor is concentrated around $\partial X_{K}\times[0,1]$.
Finally we obtain the (chiral, see the embdding (\ref{eq:embedding-spinor-3D-4D}))
fermion field $\Phi$ as source term which is directly related to
the exotic smoothness structure.

\subsection{Small exotic $\mathbb{R}^{4}$}

In case of a small exotic $\mathbb{R}^{4}$ 
\[
\mathbb{R}_{\theta}^{4}=int\left(A_{cork}\cup_{N}W\cup_{N}W\cup_{N}\cdots\right)
\]
we have a different decomposition (see \cite{BizGom:96} for an explicit
handle decomposition) using the machinery of Casson handles. But the
main results remain the same, i.e. we end up with the action 
\[
S_{EH}(R^{4})=\intop_{\mathbb{R}_{\theta}^{4}}R\sqrt{g}d^{4}x+\intop_{\mathbb{R}_{\theta}^{4}}\overline{\Phi}D\Phi\sqrt{g}d^{4}x
\]
but with an important difference. The spinor $\Phi$ is concentrated
along the boundary regions $N\times[0,1]$ like in the previous case
but now the underlying structure of the decomposition is a tree (the
tree of the Casson handle). From the physical point of view, we obtain
the creation of spinors if we go along this tree.

\section{Conclusion}

In this paper we confirmed the Brans conjecture in the form that exotic
smoothness is a generator of sources in gravity. As example we choose
the exotic $\mathbb{R}^{4}$ but the proof is general enough to include
also all other cases. The compact case was confirmed in \cite{AsselmeyerRose2012}.
As a technical tool we used the spin representation of immersed surfaces
to describe fermions as knot complements. It is interesting that fermions
are created naturally in both families (large and small) of exotic
$\mathbb{R}^{4}$'s. By using more complicated knots, one can also
descibe the interaction between the fermions (see \cite{AsselmeyerRose2012}
again). These connecting pieces are so-called torus bundles (remember
the boundary of the knot complement is a torus). There are three types
of trous bundles and we related them to the known gauge theories.
In our forthcoming work, we will describe this relation more fully.
Secondly we have done a lot of work to show a relation to quantum
gravity.

\section*{Acknowledgement}

This work was partly supported (T.A.) by the LASPACE grant. The authors
acknowledged for all mathematical discussions with Duane Randall,
Robert Gompf and Terry Lawson.

\appendix

\section{Connected and boundary-connected sum of manifolds\label{sec:Connected-and-boundary-connected}}

Now we will define the connected sum $\#$ and the boundary connected
sum $\natural$ of manifolds. Let $M,N$ be two $n$-manifolds with
boundaries $\partial M,\partial N$. The \emph{connected sum} $M\#N$
is the procedure of cutting out a disk $D^{n}$ from the interior
$int(M)\setminus D^{n}$ and $int(N)\setminus D^{n}$ with the boundaries
$S^{n-1}\sqcup\partial M$ and $S^{n-1}\sqcup\partial N$, respectively,
and gluing them together along the common boundary component $S^{n-1}$.
The boundary $\partial(M\#N)=\partial M\sqcup\partial N$ is the disjoint
sum of the boundaries $\partial M,\partial N$. The \emph{boundary
connected sum} $M\natural N$ is the procedure of cutting out a disk
$D^{n-1}$ from the boundary $\partial M\setminus D^{n-1}$ and $\partial N\setminus D^{n-1}$
and gluing them together along $S^{n-2}$ of the boundary. Then the
boundary of this sum $M\natural N$ is the connected sum $\partial(M\natural N)=\partial M\#\partial N$
of the boundaries $\partial M,\partial N$.

\section{Spin $\frac{1}{2}$ from space a la Friedman and Sorkin\label{sec:Spin-from-space}}

As shown by Friedman and Sorkin \cite{FriedmanSorkin1980}, the calculation
of the angular momentum in the ADM formalism is connected to special
diffeomorphisms $R(\theta)$ (rotation parallel to the boundary w.r.t.
the angle $\theta$). So, one can speak of spin $\frac{1}{2}$, in
case of $R(2\pi)\not=-1$. Interestingly, all hyperbolic 3-manifolds
having these diffeomorphisms.

In the following we made use of the work \cite{FriedmanSorkin1980}
in the definition of the angular momentum in ADM formalism. In this
formalism, one has the 3-manifold $\Sigma$ together with a time-like
foliation of the 4-manifold $\Sigma\times\mathbb{R}$. For simplicity,
we consider the interior of the 3-manifold or we assume a 3-manifold
without boundary. The configuration space $\mathcal{M}$ in the ADM
formalism is the space of all Riemannian metrcs of $\Sigma$ modulo
diffeomorphisms. On this space we define the linear functional $\psi:\mathcal{M}\to\mathbb{C}$
calling it a state. In case of a many-component object like a spinor
one has the state $\psi:\mathcal{M}\to\mathbb{C}^{n}$. Let $g_{ab}$
be a metric on $\Sigma$ and we define the generalized position operator
\[
\hat{g}_{ab}\psi(g)=g_{ab}\psi(g)
\]
together with the conjugated momentum 
\[
\hat{\pi}^{ab}\psi(g)=-i\frac{\delta}{\delta g_{ab}}\psi(g)\quad.
\]
Let $\phi_{\alpha}$ with $\alpha=1,2,3$ be vector fields fulfilling
the commutator rules $[\phi_{\alpha},\phi_{\beta}]=-\epsilon_{\alpha\beta\gamma}\phi_{\gamma}$
generating an isometric realization of the $SO(3)$ group on the 3-manifold
$\Sigma$. The angular momentum corresponding to the initial point
$(g_{ab},\pi^{ab})$ with the conjugated momentum $\pi^{ab}=(16\pi)^{-1}(-K^{ab}+g^{ab}K)\sqrt{g}$
(in the ADM formalism) and the extrinsic curvature $K_{ab}$ is given
by 
\[
J_{\alpha}=-\int\limits _{\Sigma}\mathcal{L}_{\phi_{\alpha}}(g_{ab})\pi^{ab}\: d^{3}x
\]
with the Lie derivative $\mathcal{L}_{\phi_{\alpha}}$ along $\phi_{\alpha}$.
The action of the corresponding operator $\hat{J}_{\alpha}$ on the
state $\psi(g)$ can be calculated to be 
\[
\hat{J}_{\alpha}\psi(g)=-i\frac{d}{d\theta}\psi\circ R_{\alpha}(\theta)^{*}(g)|_{\theta=0}
\]
where $R_{\alpha}(\theta)$ is a 1-parameter subgroup of diffeomorphisms
generated by $\phi_{\alpha}$. Then a rotation will be generated by
\[
\exp(2\pi i\hat{J})\psi=\psi\circ R(2\pi)^{*}\qquad.
\]
Now a state $\psi$ carries spin $\frac{1}{2}$ iff $\psi\circ R(2\pi)^{*}=-\psi$
or $R(2\pi)=-1$. In this case the diffoemorphism $R(2\pi)$ is not
located in the component of the diffeomorphism group which is connected
to the identity (or equally it is not generated by coordinate transformations).

\section{Scalar curvature and energy density\label{sec:Scalar-curvature-and-energy-density}}

Let us consider a Friedmann-Robertson-Walker-metric 
\[
ds^{2}=dt^{2}-a(t)^{2}h_{ik}dx^{i}dx^{k}
\]
on $N\times[0,1]$ with metric $h_{ik}$ on $N$ and the Friedmann
equation

\[
\left(\frac{\dot{a}(t)}{c\cdot a(t)}\right)^{2}+\frac{k}{a(t)^{2}}=\kappa\frac{\rho}{3}
\]
with the scaling factor $a(t)$, curvature $k=0,\pm1$ and $\kappa=\frac{8\pi G}{c^{2}}$.
As an example we consider a 3-dimensional submanifold $N$ with energy
density $\rho_{N}$ and curvature $R_{N}$ (related to $h$) fixed
embedded in the spacetime. Next we assume that the 3-manifold $N$
posses a homogenous metric of constant curvature. For a fixed time
$t$, the scalar curvature of $N$ is proportional to 
\[
R_{N}\sim\frac{3k}{a(t)^{2}}
\]
and by using the Friedmann equation above, one obtains 
\[
\rho_{N}=\frac{1}{\kappa}R_{N}+\rho_{c}
\]
with the critical density 
\[
\rho_{c}=\frac{3}{\kappa}\left(\frac{\dot{a}(t)}{c\cdot a(t)}\right)^{2}=\frac{3H^{2}}{\kappa}
\]
and the Hubble constant $H$ 
\[
H=\frac{\dot{a}}{c\cdot a}\quad.
\]
The total energy of $N$ is given by 
\begin{equation}
E_{N}=\intop_{N}\rho_{N}\,\sqrt{h}d^{3}x=\frac{1}{\kappa}\intop_{N}R_{N}\sqrt{h}d^{3}x+\rho_{c}\cdot vol(N)\,.\label{eq:total-energy}
\end{equation}
For a space with constant curvature $R_{N}$ we obtain 
\begin{equation}
E_{N}=\left(\frac{1}{\kappa}R_{N}+\rho_{c}\right)\cdot vol(N)\label{eq:total-energy-constant-curvature}
\end{equation}
 
\end{document}